\documentclass[aps,prb,preprint,superscriptaddress,floatfix]{revtex4}
\usepackage{graphicx}
\usepackage{amssymb}
\usepackage{amsmath}
\usepackage{dcolumn}
\usepackage{color}
\begin{document}

\title{Magnetic Ordering in the Frustrated Heisenberg Chain System Cupric Chloride, CuCl$_2$}

\author{M. G. Banks}
\affiliation{Max Planck Institut f\"ur Festk\"orperforschung,
Heisenbergstr. 1, D-70569 Stuttgart, Germany}

\author{R. K. Kremer}
\affiliation{Max Planck Institut f\"ur Festk\"orperforschung,
Heisenbergstr. 1, D-70569 Stuttgart, Germany}

\author{C. Hoch}
\affiliation{Max Planck Institut f\"ur Festk\"orperforschung,
Heisenbergstr. 1, D-70569 Stuttgart, Germany}

\author{A. Simon}
\affiliation{Max Planck Institut f\"ur Festk\"orperforschung,
Heisenbergstr. 1, D-70569 Stuttgart, Germany}

\author{B. Ouladdiaf}
\affiliation{Institut Laue-Langevin, B.P. 156, 38043 Grenoble,
France}

\author{J.-M. Broto}
\affiliation{Laboratoire National des Champs  Magn$\rm{\acute{e}}$tiques Puls$\rm{\acute{e}}$s - 31432 Toulouse, France}

\author{H. Rakoto}
\affiliation{Laboratoire National des Champs Magn$\rm{\acute{e}}$tiques Puls$\rm{\acute{e}}$s - 31432 Toulouse, France}

\author{C. Lee}
\affiliation{Department of Chemistry, North Carolina State
University, Raleigh, North Carolina 27695-8204, U.S.A.}

\author{M.-H. Whangbo}
\affiliation{Department of Chemistry, North Carolina State
University, Raleigh, North Carolina 27695-8204, U.S.A.}

\date{\today}

\begin{abstract}
We report a detailed examination the magnetic structure of anhydrous cupric chloride CuCl$_2$
 carried out by powder neutron diffraction, magnetic susceptibility and specific heat measurements on polycrystalline and single crystal samples
as well as an evaluation of the spin exchange interactions by first principles  density
functional theory (DFT) calculations.
Anhydrous CuCl$_2$ shows one dimensional antiferromagnetic behavior and long range antiferromagnetic ordering below a N$\acute{\rm e}$el temperature of 23.9 K. Neutron powder and single crystal diffraction reveal that, below  23.9 K, CuCl$_2$ undergoes a phase transition  into
an incommensurate magnetic structure (propagation vector (1,0.2257,0.5) with a spin-spiral propagating along $b$ and the moments confined in the  $bc$ crystallographic plane. Our  DFT calculations show that  the spin-spiral results from competing ferromagnetic nearest neighbor and antiferromagnetic next-nearest neighbor spin-exchange interaction along the spin chains. Implications for possible multiferroic behavior of CuCl$_2$ are discussed.
\end{abstract}

\maketitle

\section{Introduction}

Cupric chloride CuCl$_2$ has a crystal structure in which layers of
composition CuCl$_2$ stack along the $c$-direction with van der Waals
interactions between them. Each CuCl$_2$ layer consists of CuCl$_2$ ribbons
made up of edge-sharing CuCl$_4$ square planes such that the axial positions
of each CuCl$_4$ square
are capped by the Cl atoms of its adjacent ribbons
with long Cu...Cl distances. Thus, each Cu$^{2+}$ (d$^9$, $S$ = 1/2) ion is contained
in an axially elongated CuCl$_6$ octahedron, as expected for a Jahn-Teller
active Cu$^{2+}$ ion.
The magnetic susceptibility measurements exhibit a broad maximum,
characteristic of a short-range ordering predicted for low-dimensional magnetic systems.\cite{DeHaas1931,Starr1940,Stout1962,Barraclough1964}
Historically, CuCl$_2$ was among the very first magnetic systems whose
bulk magnetic properties were analyzed in terms of a linear spin chain model.
Heat capacity and NQR measurements indicated
a N$\acute{\rm e}$el temperature $T_{\rm N} \sim$ 23.9 K for CuCl$_2$.
To the best of our knowledge the
magnetic structure of CuCl$_2$ has remained unresolved until
now.

CuCl$_2$ is a chemically simple one-dimensional (1D) AFM spin 1/2 quantum
chain system. However, by analogy to the CuO$_2$ ribbon chains
found in the magnetic oxides such as LiCuVO$_4$ and LiCu$_2$O$_2$,
the next-nearest-neighbor (NNN) spin exchange coupling
along the chain may be essential in addition to the nearest-neighbor (NN) spin exchange
in understanding the magnetic properties of CuCl$_2$.
For these CuO$_2$ chains, made up of edge-sharing CuO$_4$ square planes,
the NN spin exchange is ferromagnetic (FM),
while the NNN spin exchange is strongly AFM,
and the resulting geometric spin frustration gives rise to
an incommensurate spiral spin ordering and ferroelectric polarization.
Similarly, CuCl$_2$ might undergo a spiral spin ordering within
each Cu chain and a long-range N$\acute{\rm e}$el ordering as do LiCuVO$_4$ and LiCu$_2$O$_2$.
In the present work we examine the magnetic structure of
CuCl$_2$ in detail by powder neutron diffraction, magnetic susceptibility
and specific heat measurements as well as
by evaluating its spin exchange interactions on the basis of
first principles density functional DFT calculations.

\section{Experimental}

Single crystals of CuCl$_2$ were grown  in quartz glass ampoules by the Bridgman technique using commercially available
powder (Alfa Aesar, ultra dry, purity 99.995\%). Special care was taken in the design of the
ampoules to support the elevated vapor pressure of chlorine above
the melting point of CuCl$_2$. Well-crystallized needle-shape crystals,
obtained with the needle axis oriented along the crystallographic
$b$ axis were found to be twinned   in the $a$$c$-plane. Due to the moisture sensitivity of
CuCl$_2$ all handling of the samples was done in an Ar filled glove
box. Temperature and field dependent magnetizations
(0 T $< B_{\rm ext} <$ 7 T) were measured using a
superconducting quantum interference device magnetometer (MPMS,
Quantum Design, 6325 Lusk Boulevard, San Diego, CA.). Magnetizations
up to 50 T were measured in pulsed magnetic fields at the Laboratoire National des Champs
Magn$\acute{\rm e}$tiques Puls$\acute{\rm e}$s (Toulouse, France)
at temperatures of 1.4 and 4.2 K with the field aligned along the crystal $b$
axis. The heat capacities of crystals ($m \sim$ 20 mg)  were
measured using a commercial Physical Property Measurements System
calorimeter (Quantum Design, 6325 Lusk Boulevard, San Diego, CA.)
employing the relaxation method in magnetic fields up to 9 T aligned
parallel and perpendicular to the crystal $b$ axis. To thermally
anchor the crystals to the heat capacity platform, a minute amount
of Apiezon N grease was used. The heat capacity of the platform and
the grease was determined in a separate run and subtracted from the
total heat capacity. The heat capacity of powder samples ($m \sim$
300 mg) was measured in a home-built  fully automatic
Nernst-type adiabatic calorimeter similar to that described in
detail in Refs. \onlinecite{MBThesis,Gmelin}. The powder samples were encapsulated in
Duran glass flasks under $\sim$ 900 mbar $^4$He atmosphere to enable rapid thermal equilibration.
The flasks  were attached with a minute amount of
Apiezon N vacuum grease to the sapphire platform which carries an
 deposited thin-film stainless steel heater and a calibrated Cernox CX-1050-SD thermometer (Lake Shore Cryotronics, Inc.
575 McCorkle Blvd.
Westerville OH 43082). The heat
capacities of the sapphire sample platform, the glass flask and the Apiezon vacuum grease were
determined in separate runs and subtracted from the total heat
capacity.
EPR spectra were measured with a Bruker ER040XK microwave X-band
spectrometer and a Bruker BE25 magnet equipped with a BH15 field
controller calibrated against DPPH. A stepper-motor-controlled goniometer was used to rotate the crystals around an axis perpendicular to the
magnetic field plane. The crystals were oriented optically with a precision of $\sim \pm 5^{\rm o}$.
100 kHz field modulation  was used to record the first derivative of the signal intensity.
The spectra were fitted to a resonance curve with  Lorentzian lineshape (zero dispersion)  by varying the resonance position as well as the linewidth, intensity and background parameters.

Powder neutron diffraction patterns were collected on the
high-intensity two-axis diffractometer D20 (Institut Laue-Langevin,
Grenoble) at a wavelength of 2.4 \AA\  and 1.88 \AA\  in the
temperature range 1.8 K$ < T <$ 50K. \cite{D20} Single crystal
diffraction was carried out on the four-circle diffractometer D10
(Institut Laue-Langevin, Grenoble) at a wavelength of 2.36 \AA\  on a
crystal of $\sim$50 mm$^3.$ \cite{D10}

\section{Experimental Results}

\subsection{Crystal Structure}
\begin{figure}
\includegraphics[width=8cm,angle=0]{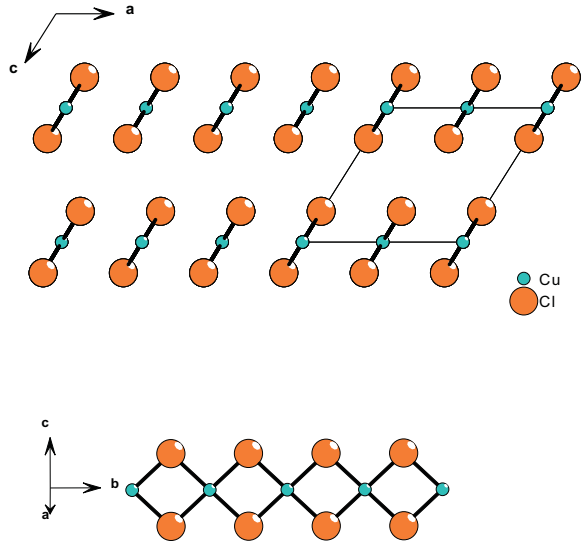}
\caption{\protect \label{structure}(Color online) Crystal structure
of CuCl$_2$ projected along [010] drawn using parameters according
to Refs. \onlinecite{Wells1947,Burns1993}. A unit cell is outlined by the
black solid lines. The Cu atoms are represented by small (blue), and the Cl
atoms by large (orange) circles.
The lower part of the figure
highlights a CuCl$_2$ ribbons with the $b$ axis along the
ribbons. The planes of the ribbons are parallel to the $bc$ plane.}
\end{figure}

At room temperature, anhydrous cupric chloride CuCl$_2$ was reported to
crystallize with a monoclinic structure
(space group $C$2/$m$) with 2 formula units per unit
cell.\cite{Wells1947,Burns1993}
The crystal structure of CuCl$_2$ consists of
CuCl$_2$ slabs parallel to the $ac$-plane that are made up of
CuCl$_2$ chains running along the crystallographic $b$ axis.
Such slabs are  interconnected along
the $c$ axis via van der Waals contacts (cf. Figure  \ref{structure}).

\begin{figure}
\includegraphics[width=10cm,angle=0]{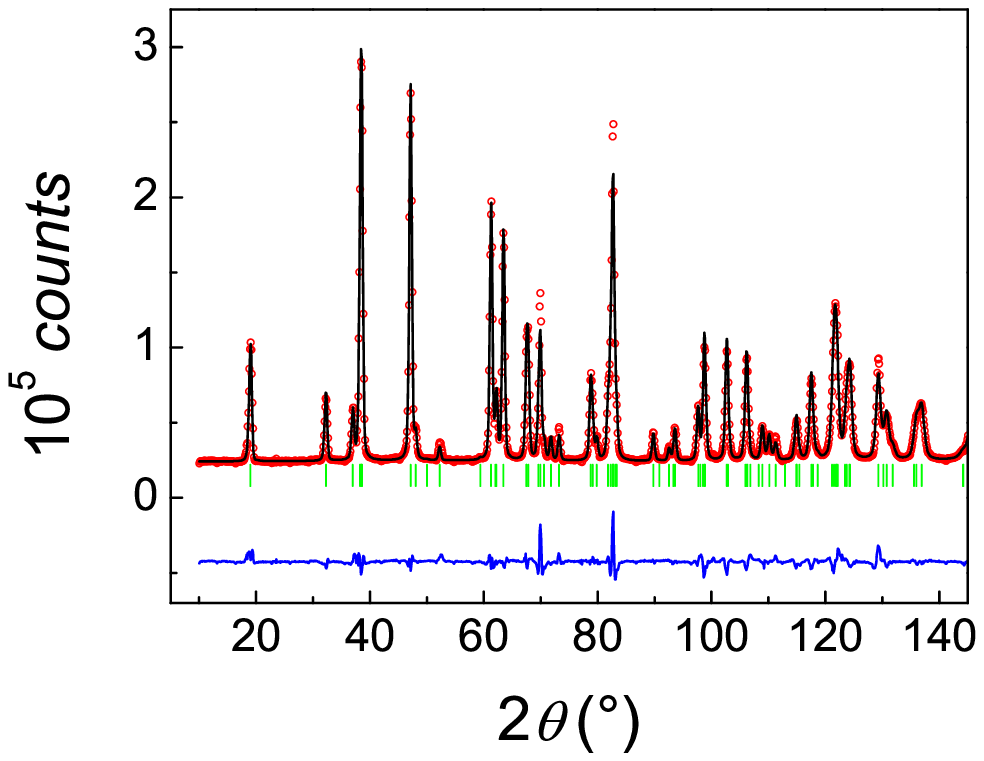}
\caption{\protect \label{nuc50}(Color online) Neutron diffraction
pattern (red $\circ$) ($\lambda$=1.888 \AA\  ) of a powder sample of
CuCl$_2$ collected at 42.9 K at the high intensity diffractometer D20
plotted in comparison with the pattern obtained from a profile
refinement (solid black line). The difference is shown in
the lower part by the (blue) solid line. The positions of the
Bragg reflections used to simulate the patterns are indicated by the
small (green) vertical bars in the lower part of the figure.}
\end{figure}

To confirm the crystal structure details of CuCl$_2$ and to search for structural phase
transitions at low temperatures, we
collected a series of neutron diffraction patterns on the diffractometer D20 at temperatures
2 K $ < T <$ 50K.
Patterns with sufficient intensity but improved resolution were collected using neutrons with a
wavelength of $\lambda$ = 1.88 \AA\  (Ge monochromator)
suitable for Rietveld profile refinements. These were done with the FullProf program package\cite{FullProf}.
The patterns were refined assuming
the space group $C$2/$m$ with the Cu atoms in the Wyckoff position
(2a) (0,0,0) and the Cl atoms in the position (4i) ($x$,0,$z$)
using a slightly asymmetric pseudo-Voigt profile shape to model the
Bragg reflections. Within experimental error the atom coordinates
$x$ and $z$ for the Cl atoms are independent of the temperature and
amount to $x$=0.5065(2) and $z$=0.2365(3).
Figure \ref{nuc50} displays the experimental pattern collected at 42.9 K
and the results of the Rietveld refinement ($\lambda$=1.88 \AA\ ). Apart from
changes in the lattice parameters, the structure remains unchanged
with respect to the room temperature crystal structure of CuCl$_2$
described by Wells and Burns \textit{et al.}.\cite{Wells1947,Burns1993}
The atom position
parameters show no noticeable difference to those of the room
temperature structure. The structure parameters at 2.5 and 42.9 K are
summarized in Table \ref{tabstruc}.

\begin{table}[ht]
\begin{ruledtabular}
\begin{tabular}{cccccc}
$T$(K)& $a$(\AA\ )&  $b$ (\AA\ ) & $c$ (\AA\ ) & $\beta$ ($^{\rm o}$)  &  method\\
\hline\\
2.55 & 6.7986(3) &  3.29418(12) & 6.7718(3) &  122.45(2)  & neut \\ \\
42.9 & 6.8010(3) & 3.29437(12)  & 6.7768(3) &  122.47(2)  & neut \\ \\
293  & 6.8973(4)   & 3.2961(2)    & 6.8160(4)   &  122.239(3) & x-ray  \\ \\
293  & 6.9038(9)   & 3.2995(4)    & 6.824(1)    &  122.197(8) & x-ray (Ref. \onlinecite{Burns1993})\\
\end{tabular}
\end{ruledtabular}
\caption[]{Lattice parameters  of CuCl$_2$ obtained from neutron powder and x-ray diffraction patterns (Cu K$ \alpha_1$ radiation)
at the indicated temperatures. Literature values are given for comparison.
\label{tabstruc}}%
\end{table}

\begin{figure}[ht]
\includegraphics[width=8cm]{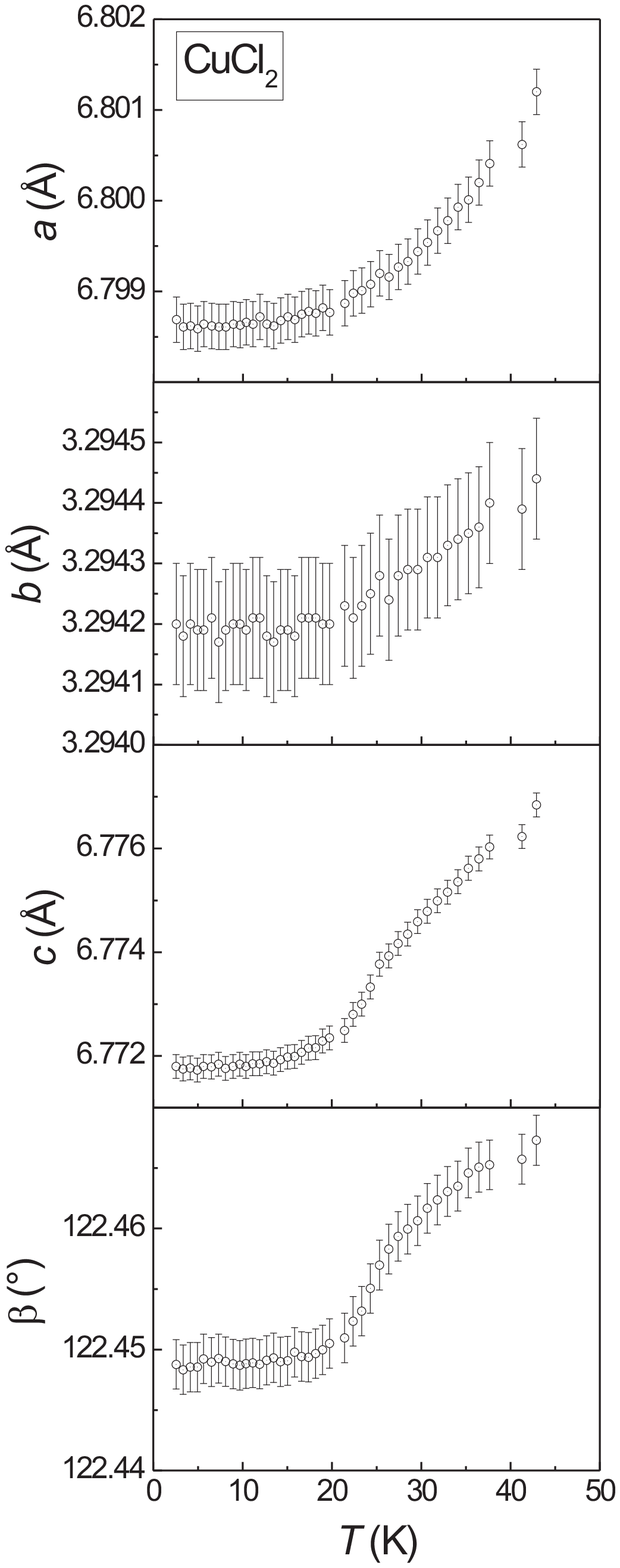}
\caption{\protect \label{latpar} Temperature dependence of the
refined lattice parameters of CuCl$_2$ resulting from Rietveld
profile refinements of the neutron powder diffraction patterns (details see text).}
\end{figure}

The temperature dependence of the lattice parameters is displayed in
Figure  \ref{latpar}. The lattice parameters show a monotonic
decrease with temperature. Below $\sim$ 15 K they level off and remain
constant to the lowest temperatures. Near $\sim$ 24 K,
especially $c$ and $\beta$, exhibit noticeable
changes in the slope, which indicate a magnetoelastic response of the
lattice to the onset of long-range AFM ordering,
as is evidenced by our magnetization and heat capacity experiments  discussed in the following paragraphs.

Single-crystal x-ray diffraction on crystals of CuCl$_2$ were carried out using a laboratory based diffractometer (PDS II, Stoe \& Cie., Darmstadt, Germany). These measurements confirmed the crystal structure determined from neutron powder diffraction and showed the crystals to be twinned in the \textit{ac}- plane. The twinning matrix \textit{T} relating the Miller indices of the twinned individua  was determined to

\begin{equation}
T=
\begin{pmatrix}
       1 &  0 & -1  \\
        0 & -1 & 0  \\
       1 &  0 & 0
       \end{pmatrix}.
\end{equation}

The twinning reverses the $b$-axis and rotates the the axes in the reciprocal space,
such that twin-related axes enclose an angle of 180$^{\rm o}$ - $\beta \approx$ 57$^{\rm o}$.

\section{Magnetic Properties}

\subsection{Magnetic Susceptibility Measurements}

\begin{figure}
\includegraphics[width=9cm]{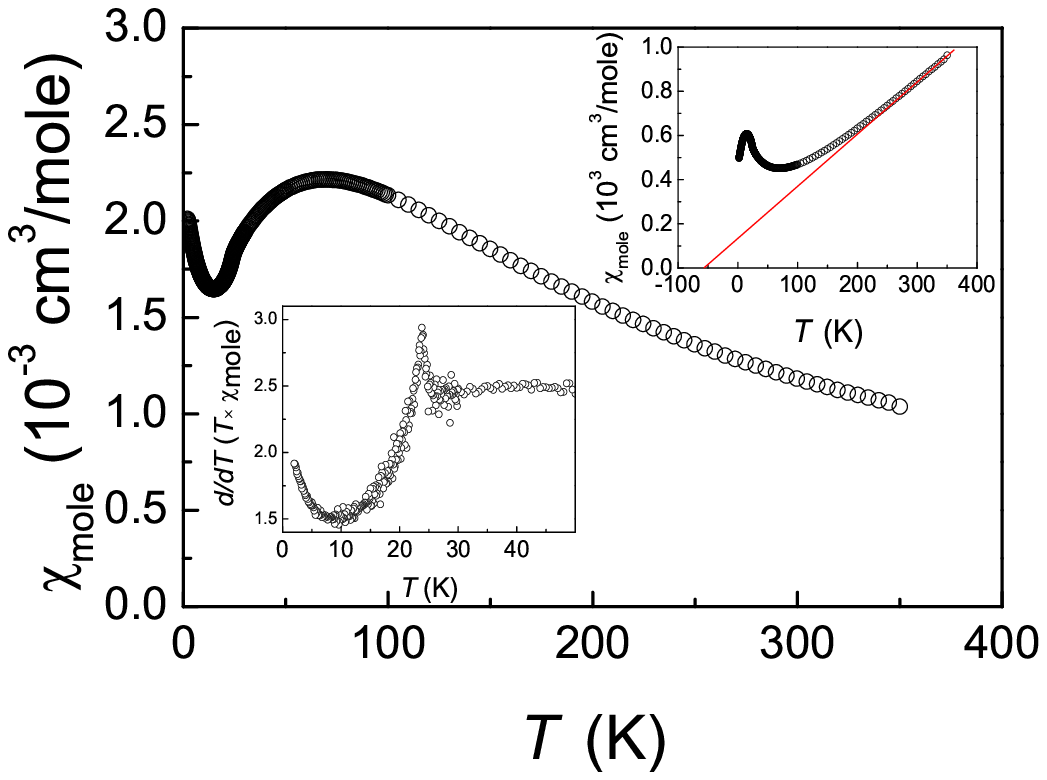}[h]
\caption{\protect \label{Chipowder}(Color online) Temperature dependence
of the magnetic susceptibility (not corrected for temperature independent diamagnetic or Van Vleck paramagnetic contributions) of a polycrystalline sample of CuCl$_2$ determined in an external magnetic field of 1 T. The upper inset shows the reciprocal susceptibility with the (red)
solid line indicating a Curie-Weiss law with an intercept on the abscissa at $\sim$ -60K.
The lower inset shows $d/dT (\chi_{\rm mol} \times T$ ('Fishers heat capacity').
A sharp $\lambda$-type anomaly at 23.9(3) K indicates the onset of long-range AFM ordering.}
\end{figure}

Figure  \ref{Chipowder} displays the magnetic susceptibility of a polycrystalline sample of CuCl$_2$.
The susceptibility is characterized by a broad maximum centered at $\sim$ 70 K and a Curie-Weiss law (Eq. (\ref{eqCW}) behavior at
high temperatures (see the detailed discussion of the susceptibility measurement on a crystal below)
with a negative Curie-Weiss temperature of $\sim$ -60 K and a powder $g$-factor (see Eq. (\ref{eqCW}) of $\sim$2.13.

The negative Curie-Weiss temperature
proves predominant AFM exchange interactions, and the broad maximum reveals AFM
short range ordering indicative of a low-dimensional magnetic behavior. A kink-like change of the slope
near $\sim$ 24 K is due to the onset of long-range AFM ordering. A $\lambda$-type anomaly is clearly
revealed by the temperature derivative of the quantity $\chi_{\rm mol} \times T$
('Fisher's heat capacity', cf. Ref. \onlinecite{Fisher1962}) shown in the inset of Figure  \ref{Chipowder}
from which the  N\'{e}el temperature $T_{\rm N}$  is  determined to be 23.9(3) K. $T_{\rm N}$
coincides with the anomalies observed in the
temperature-dependence of the lattice parameters (Figure  \ref{latpar}). When fitting a Curie-Weiss law to the
high temperature data of the powder susceptibility it is found that the fitted parameters
(the slope of the reciprocal susceptibility i.e. the effective magnetic moment, and the intercept) are strongly correlated
and depend essentially on temperature-independent contributions to the susceptibility (diamagnetic and Van Vleck susceptibilities).

\begin{figure}
\includegraphics[width=9cm]{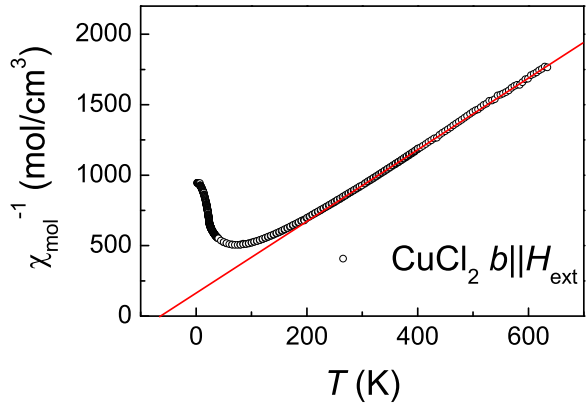}
\caption{\protect \label{Chixtalrec}(Color online) Reciprocal magnetic susceptibility (uncorrected for diamagnetic and Van Vlexck contributions)
of a crystal of CuCl$_2$ determined in an external magnetic field of 1 T oriented parallel
to the crystallographic $b$-axis.
The (red) solid line represents a Curie-Weiss law including a temperature independent term $\chi_0$ fitted to the data with parameters as given in the text.}
\end{figure}

In order to obtain a reliable fit of the magnetic susceptibility of CuCl$_2$ over a wide temperature range and as well as the temperature independent diamagnetic and paramagnetic corrections to the susceptibility we determined
the magnetic susceptibility on a crystal ($B_{\rm ext}$=1T $\parallel b$) up to 600 K. At high temperatures the magnetic susceptibility follows a  Curie-Weiss law as demonstrated by the plot of the reciprocal susceptibility in Figure \ref{Chixtalrec}.
Deviations from the Curie-Weiss law are seen already at temperature below $\sim$ 175 K.

At high temperatures, besides the spin susceptibility, $\chi_{\rm spin} (T)$, temperature independent diamagnetic orbital contributions, $\chi_{\rm dia}$, from the core electronic shells and the Van Vleck paramagnetism,
$\chi_{\rm VV}$, from the open shells of the Cu$^{2+}$ ions become important.
$\chi_{\rm dia}$ can be estimated from the increments using  Selwood's tabulated values as -12$\times$10$^{-6}$ cm$^3$/mole and -26$\times$10$^{-6}$ cm$^3$/mole for
Cu$^{2+}$ and Cl$^-$, respectively, giving a total temperature independent diamagnetic susceptibility for CuCl$_2$ of -64$\times$10$^{-6}$ cm$^3$/mole.\cite{Selwood1956}
An estimate of the Van Vleck contribution, $\chi_{\rm VV}$ can be obtained from (cf. Ref \onlinecite{Lueken1999}).

\begin{equation}
\label{chiVV}
\chi_{\rm TIP} \approx \frac{4 N_{\rm A} \mu_{B}^2}{\Delta},
\end{equation}

where $N_{\rm A}$ is Avogadro's constant and $\Delta$ = 10$|D\,q|$ is the energy separation of the $d$ orbital states in a cubic crystal field, which amounts to approximately 25$\times$10$^{3}$ cm$^{-1}$. With $N_{\rm A} \mu_B^2$/$k_B$ $\approx$ 0.375 cm$^3$ K/mole one estimates a Van Vleck susceptibility of Cu$^{2+}$ in a cubic crystal field of the
order of +43$\times$10$^{-6}$ cm$^3$/mole.
In CuCl$_2$ due to the Jahn-Teller distortion the symmetry is reduced and the Van Vleck contribution depends on the actual matrix elements and the energy separation
of the orbital states to the $d_{x^2-y^2}$ state. For a detailed discussion see for example Ref. \onlinecite{Takigawa1989}. Since these energy splittings are not known at present, we used an alternative method to determine the temperature independent contributions $\chi_0$:
In order to perform this analysis,  we assume that the magnetic susceptibility of CuCl$_2$ can be described by a sum of the temperature dependent spin susceptibility $\chi_{\rm spin}(T)$ and a temperature independent contribution $\chi_0$

\begin{equation}
\label{eqchitot}
\chi (T) =  \chi_{\rm 0} + \chi_{\rm spin}(T)
\end{equation}

where $\chi_0$ is composed by the sum of the diamagnetic and the Van Vleck contribution according to

\begin{equation}
\label{eqchi0}
\chi_0 = \chi_{\rm dia} + \chi_{\rm VV}.
\end{equation}

At sufficiently high temperatures, $T >> \theta$, the spin susceptibility $\chi_{spin}$(T) can be approximated by a Curie-Weiss law according to

\begin{equation}
\label{eqCW}
\chi_{\rm spin} \approx \chi_{\rm CW} = \frac{N_{\rm A} g^2 \mu_{B}^2 S(S+1)}{3 k _{B} (T - \theta)}.
\end{equation}

When fitting the high temperature susceptibility to
the sum of equations (\ref{eqchi0}) and (\ref{eqCW})
it is found that the fitting parameters,
$g$, $\chi_0$ and $\theta$, are strongly correlated, and meaningful results cannot be obtained without fixing one of them.
In order to reduce the number of parameters we therefore determined the g-factor, $g_b$ independently by an EPR experiment on a crystal (see below)
to $g_b$ = 2.050(1). The EPR experiment also revealed that our CuCl$_2$ crystals consist of mirror twins in the $ab$
crystallographic plane with the mirror twins having the same $b$ axis.
By using $g_b$ = 2.05 and fitting Eq. (\ref{eqchitot}) to
the (uncorrected) susceptibility data for $T > 150$ (see Figure  \ref{Chixtalrec}), we determine the following parameters:

\begin{equation*}
\label{fitresults1}
\chi_0 = -18(1)\times 10^{-6} {\rm cm^3/mole},
\end{equation*}
\begin{equation*}
\label{fitresults2}
\theta = -71(1) {\rm K}.
\end{equation*}

\subsection{Magnetization Measurements}

\begin{figure}
\includegraphics[width=8cm,angle=0]{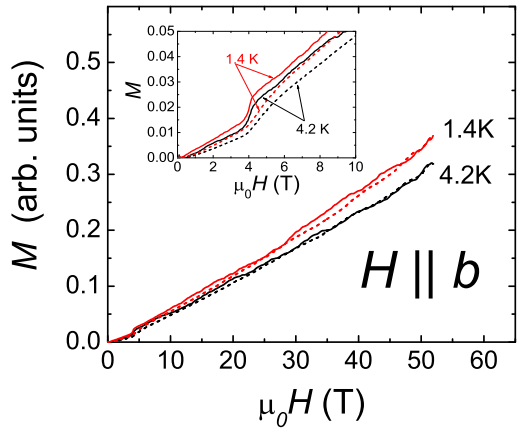}
\caption{\protect \label{Magpulse}(Color online) Magnetization of a
crystal of CuCl$_2$ measured at 1.4 K and 4.2 K in a pulsed magnetic field with the
magnetic field aligned along $b$. The inset shows a magnification of
the low field data. The solid and dashed lines represent data at
1.4 K and 4.2 K, respectively, measured with increasing (solid line) and with decreasing (dashed line) magnetic field.}
\end{figure}

Laboratory based magnetization experiments at 1.8 K with $B_{\rm ext} \parallel b$ evidenced
a spin-flop transition near 4 T. Measurements in a pulsed magnetic field at 1.4 K up to $\sim$ 50 T
confirmed this result. Evidence for further magnetic phase transitions at fields larger than $\sim$ 4 T could  could not be found.
Figure \ref{Magpulse} displays the magnetizations measured in the pulsed field experiment.
Saturation is not seen up to 50 T.
The 1.4 K magnetization trace shows a slight upward deviation
from nonlinearity while the 4.2 K traces within experimental error
show a linear dependence of the magnetization  over the full field range.

\begin{figure}
\includegraphics[width=8cm,angle=0]{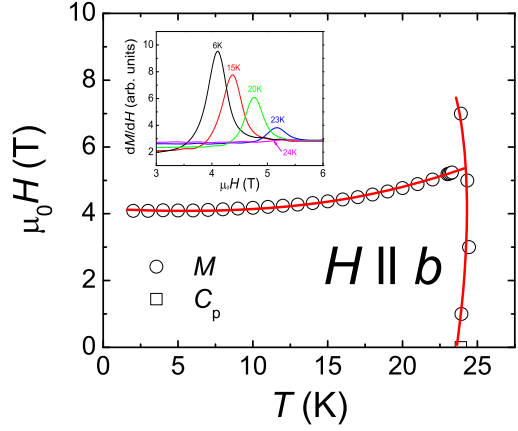}
\caption{\protect \label{MagPD}(Color online) Magnetic phase
diagram of CuCl$_2$ derived from magnetization and heat capacity
measurements with the magnetic field aligned along $b$. The inset
displays the derivative $dM$/$dH$ for selected fields $H$ from which
the phase boundaries have been extracted.}
\end{figure}

Magnetic susceptibility experiments in various magnetic fields reveal
that $T_{\rm N}$ depends only weakly
on the field. The combination of the magnetization and the
magnetic susceptibility experiments
allows us to construct a magnetic phase diagram for $B_{\rm ext} \parallel b$
as displayed in Figure
\ref{MagPD} with a bicritical point at 5.4(1) T and 24.2(1) K where
the paramagnetic, spin flop and N\'{e}el phases coexist.

\subsection{Electron Paramagnetic Resonance Measurements}

\begin{figure}
\includegraphics[width=8cm,angle=0]{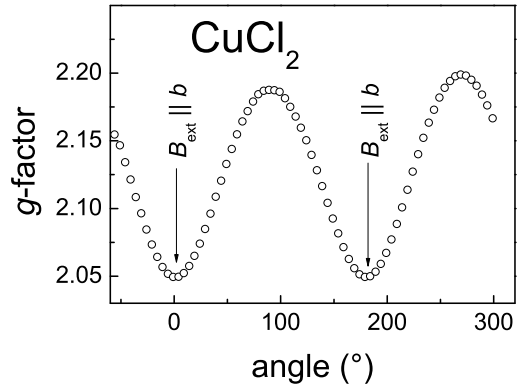}
\caption{\protect \label{gbaxis}(Color online) Angular dependence of the $g$-factor of the EPR resonance line of Cu$^{2+}$
in a crystal of CuCl$_2$ when the crystallographic $b$-axis lies within the magnetic field plane.
With $b \parallel B_{\rm ext}$ the g-factor amounts to 2.050(1).}
\end{figure}

In order to determine the $g$-factors along the principal axes of the $g$-tensor we carried
out  EPR measurements on small crystals of CuCl$_2$ at room temperature at
a microwave frequency of 9.46 GHz (X-band) (resonance field $\sim$ 0.33 T).
In a monoclinic system,  one principal axis of the $g$-tensor is fixed by symmetry and
oriented parallel to the twofold axis,
i.e. the crystallographic $b$ axis for CuCl$_2$. The remaining two principal axes of the
$g$-tensor lie within the $ac$ plane.
To obtain the $g$-factor along $b$, a crystal was oriented with the $b$-axis within the magnetic field plane. Rotation was carried out around an arbitrary axis perpendicular to $b$. A single resonance line is observed. Figure \ref{gbaxis} shows the angular dependence of the $g$-factor obtained from fits of a field derivative of a Lorentzian to the resonance line. The angular dependence of the $g-$factor shows the typical oscillating behavior. The minima occur when the field is oriented parallel to the $b$ axis. The $g$-factor with field oriented parallel to the $b$-axis amounts to

\begin{equation*}
g_b = 2.050(1).
\end{equation*}

The linewidths of the resonance lines exhibit angular dependence and they vary between 20 mT and 40 mT due to small anisotropic exchange components.\cite{MBThesis} Details will be discussed in a forthcoming paper. \cite{Rekretbp}

\begin{figure}
\includegraphics[width=8cm,angle=0]{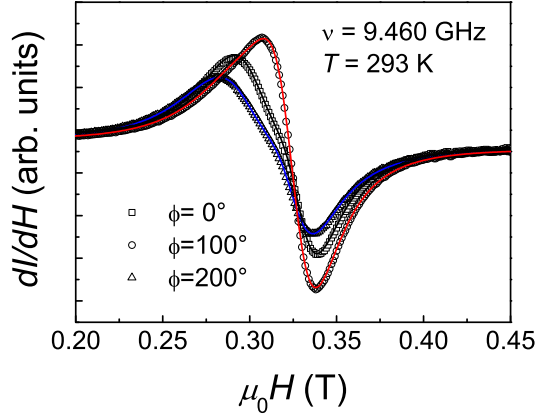}
\caption{\protect \label{EPR3lines}(Color online) Symbols: EPR spectra of CuCl$_2$ with the external magnetic
field oriented in the crystallographic $ac$-plane at (arbitrary) angles as given in the inset. The solid
lines are fits to the measured spectra assuming two overlapping derivatives of Lorentzian resonance lines with separate
resonance fields, linewidths and intensities, respectively.}
\end{figure}

If the magnetic field was oriented in the $ac$ plane of the crystal (rotation around $b$-axis), we observed
an EPR spectrum that consists of two overlapping resonance lines,  indicating that the crystals were twinned.
The resonance fields and the linewidths of each line and the intensity were obtained by fitting to the spectra
the sum of two Lorentzian absorption lines (see Figure \ref{EPR3lines}).

\begin{figure}
\includegraphics[width=8cm,angle=0]{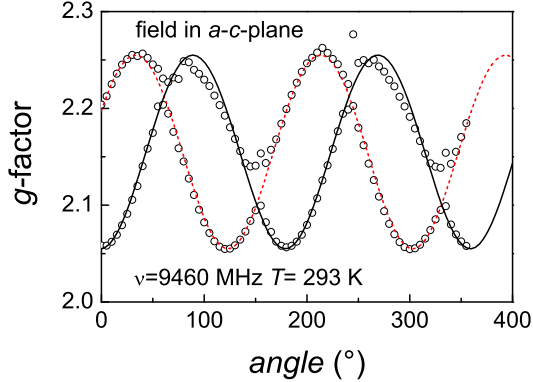}
\caption{\protect \label{EPRacplane}(Color online) Symbols: EPR spectra of CuCl$_2$ with the external magnetic
field oriented in the crystallographic $ac$-plane at (arbitrary) angles as given in the inset. The solid
lines are fits to the measured spectra assuming two overlapping Lorentzian resonance lines with individual
resonance fields, linewidths and intensities each.}
\end{figure}

The  $g$-factors of the two lines are shown in Figure \ref{EPRacplane}.
One observes the typical oscillating angular dependence of the $g$-factors on the angle between magnetic field the crystal orientation. Both lines show the same oscillating angular dependence of
the $g$-factor, however with respect to each other shifted by $\approx$57$^{\rm o}$. The extremal values are identical within experimental errors. The intensities for the two crystallites are about the same indicating that
the investigated twinned crystallites have about the same volume.
At angles  where both twin-individua have the same resonance field and $g$-factors,
it is  difficult to deconvolute the two resonance line unambiguously. This leads to the rather high error level in the fitted parameters
especially at angles where the two $g$-factor curves intersect.

The principal $g$-factors, $g_1$ and $g_2$ were determined by comparing the results
with the general angular dependence of the $g$-factor for an axial system  (Ref. \onlinecite{Abragam1970})

\begin{equation}
\label{gfactor}
g^2(\phi) = g_1^2 \cos^2(\phi-\phi_0) + g_2^2 \sin^2(\phi-\phi_0),
\end{equation}

where $\phi$ is the angle between the external field and the direction of the principal axis $1$.
$\phi_0$ is a phase factor adjusted individually for each individual crystal.

The solid lines in Figure  \ref{EPRacplane} were calculated according to Eq. (\ref{gfactor}) using

\begin{equation*}
g_1 = 2.255(2),
\end{equation*}
\begin{equation*}
g_2 = 2.055(2).
\end{equation*}

With $g_b$ = 2.050(1), these values result in an
average $g$-factor of

\begin{equation}
\bar{g} = \sqrt{\frac{1}{3} \sum_{i} g_i^2}
\end{equation}

\begin{equation*}
\bar{g} = 2.120(5).
\end{equation*}

 This valueis in good agreement with the $g$-factor obtained from the powder susceptibility measurement (see. Figure  \ref{Chipowder}).

$g_2$ and $g_b$ are almost identical indicating that the local symmetry for Cu$^{2+}$ in CuCl$_2$ is very close to axial with
$g_1 \equiv g_{\parallel}$ =2.255 and $g_b \approx g_2 \equiv g_{\perp} \approx$ 2.05. Here, we have identified the large $g_1$ with the $g$-factor along the quasi-fourfold symmetry axis, perpendicular to the CuCl$_4$ squares in the ribbons and $g_b$ and $g_2$ with the $g$-factors for the magnetic field aligned within the planes of the ribbon.

This finding reflects the local symmetry, which is  very close to tetragonal
given the elongated CuCl$_6$ octahedron containing each Cu$^{2+}$ ion.
The four Cl atoms in the equatorial plane are all at the same distance of  2.253 \AA\ and the apical Cl atoms at a distance of 2.911~\AA\ ($T$ = 42.9 K);
the normal of the equatorial plane and the direction to the apical Cl atoms encompass an angle of only 3.2$^{\rm o}$.

The deviation of the $g$-factors of the ground state of a 3$d^9$ configuration in octahedral field with tetragonal splitting from the $g$-factor of the free electron is determined by the spin-orbit coupling parameter $\lambda$ = -$\zeta$/2$S$, where $\zeta$ is the one-electron spin-orbit coupling parameter and $S$=1/2. For Cu$^{2+}$ $\zeta$ amounts to 829 cm$^{-1}$ (Ref. \onlinecite{Abragam1970}) and the deviations from the free electron $g$-factor are given by

\begin{eqnarray}
\label{gfactors}
g_{\parallel} - 2 & = &\frac{8\lambda}{\Delta_0}\nonumber\\
g_{\perp} - 2 & = & \frac{2\lambda}{\Delta_1},\nonumber\\
\end{eqnarray}

where $\Delta_0$ and $\Delta_1$ are the energy separations between the $d_{x^2-y^2}$ state and the $d_{xy}$ and the $d_{yz,xz}$ electronic states, respectively. \cite{Abragam1970}
Using our experimental results, $\Delta_0$ and $\Delta_1$ amount to $\sim$ 26$\times$10$^{3}$ cm$^{-1}$ and $\sim$ 30$\times$10$^{3}$ cm$^{-1}$, respectively, consistent with the estimate we made from the magnitude of the temperature independent Van Vleck contribution.

The principal axes of the two crystallites are shifted by 180$^{\rm o}$ - $\beta  \approx$ 57$^{\rm o}$  with respect to each other,  where $\beta$ is the monoclinic angle (cf. Table \ref{tabstruc}). This finding  indicates that CuCl$_2$ crystals generally are twinned with the twins  related to each other by a mirror symmetry operation with the mirror symmetry plane in the $ac$ plane making the crystals to appear pseudo-hexagonal. This finding becomes essential for the assignment of the propagation vector of the magnetic structure  from elastic neutron diffraction on  CuCl$_2$ crystals.

\section{Thermal Properties}

\subsection{Heat Capacity Measurements}

\begin{figure}[h]
\includegraphics[width=9cm]{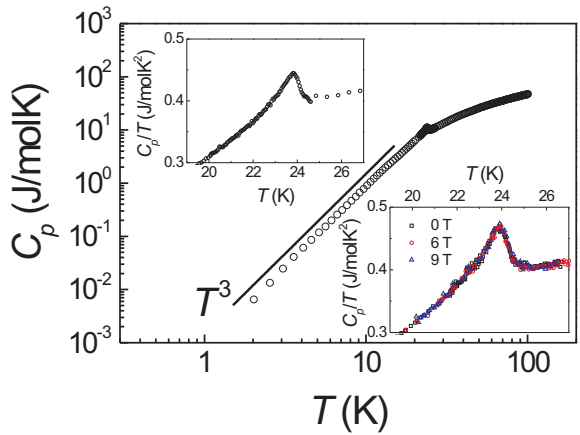}
\caption{\protect \label{heat}(Color online) Temperature dependence
of the specific heat  of CuCl$_2$. The main frame shows the heat capacity of a crystal, the lower
inset shows the  magnetic field dependence of the anomaly at $T_{\rm N}$ = 23.85(3) K. The magnetic field was
directed perpendicular to the $b$ axis. The upper inset displays the
heat capacity near $T_{\rm N}$ of a polycrystalline sample.}
\end{figure}

The heat capacity of anhydrous CuCl$_2$ has been determined before by Stout and Chisholm (Ref. \onlinecite{Stout1962}) and by Billerey \textit{et al.} (Ref. \onlinecite{Billerey1982}).
Figure \ref{heat} shows our heat capacities $C_p$ of a CuCl$_2$ crystal
and a powder sample. Our data agree very well with the data published by Stout \textit{et al.} and those of Billerey and collaborators.
The data of Stout and Chisholm are shown in Figure  \ref{heat} for comparison.
The heat capacities of our powder sample and the crystal are very
similar, and they exhibit $\lambda$-shaped anomalies due to the onset of
long-range AFM ordering. The anomalies are centered at
$\sim$23.8 K (crystal: 23.87 K; powder 23.82 K) in good agreement with the values reported by Stout \textit{et al.} (23.91$\pm$0.1 K) and Billerey \textit{et al.} (23.9$\pm$0.1 K).  The maximum value in the $\lambda$-anomaly is somewhat larger than the value reported by Stout \textit{et al.} and Billerey \textit{et al.} possibly indicating that our samples structurally are better ordered and contain less structural defects. The $\lambda$-anomaly does not change its shape nor shifts the position when a magnetic field of up to 9 T is applied perpendicular to the crystal $b$ axis.
Sufficiently below the anomaly where the lattice contribution can be
approximated by a Debye-$T^3$ power law, the total  heat capacities
follow well the same power law $C_p$ = $C_{\rm ph}$+$C_{\rm
magnon}\propto T^3$ indicating  magnon contributions from a
three-dimensional antiferromagnet.

The entropy
removed by long-range ordering signaled by the anomaly at 23.8 K amounts to 0.60(5) J/molK equivalent to
10\% of $R$ln2 as expected for a $S$=$\frac{1}{2}$ spin system.
The remaining 90\% of the entropy apparently are removed by short-range
ordering above $T_{\rm N}$.

The total heat capacity, $C_p$, of CuCl$_2$ contains  the spin (magnetic) and contributions from the phonon system (lattice contribution) according to

\begin{equation}
C_p(T) = C_{\rm mag}(T) + C_{\rm latt}(T)
\end{equation}

In order to separate the magnetic  contributions, $C_{\rm mag}$,
from the total heat capacity of CuCl$_2$, one has to determine the lattice heat contributions, $C_{\rm latt}$.  The lattice contributions can be estimated by fitting a power series expansion to the high temperature regime, where magnetic contributions are negligible, and extending the series to low temperatures. A significantly improved estimate of the lattice heat capacity is usually achieved by measuring the heat capacities of  isostructural nonmagnetic compounds.  The lattice reference systems appropriate
for CuCl$_2$  would be MgCl$_2$ and CdCl$_2$. Both compounds crystallize with a layered crystal structure similar to that of CuCl$_2$.\cite{Partin1991}  However,  the MCl$_6$ (M=Mg, Cd) octahedra of MgCl$_2$ and CdCl$_2$ do not exhibit a Jahn-Teller distortion,  and consequently the two-valent anions are coordinated regularly by six Cl atoms at the same distance of 2.64 \AA\  and 2.96 \AA\  for CdCl$_2$ and MgCl$_2$, respectively.\cite{Partin1991}.

Billerey \textit{et al.} determined the heat capacities of MgCl$_2$  in the temperature range 2 K $ < T < $ 100 K and  constructed the lattice contribution to the heat capacity of CuCl$_2$.
We  re-measured the heat capacities of a polycrystalline sample and of crystalline pieces of high purity (Alfa Aesar 99.99\%) MgCl$_2$ in an extended temperature range.  We found that our data for MgCl$_2$ obtained on two independent samples with two different experimental setups, especially, in the  temperature regime relevant for the magnetic part below 100 K deviate markedly from Billerey's data. Our values being up to 10\% smaller then Billerey's results.

\begin{figure}
\includegraphics[width=9cm]{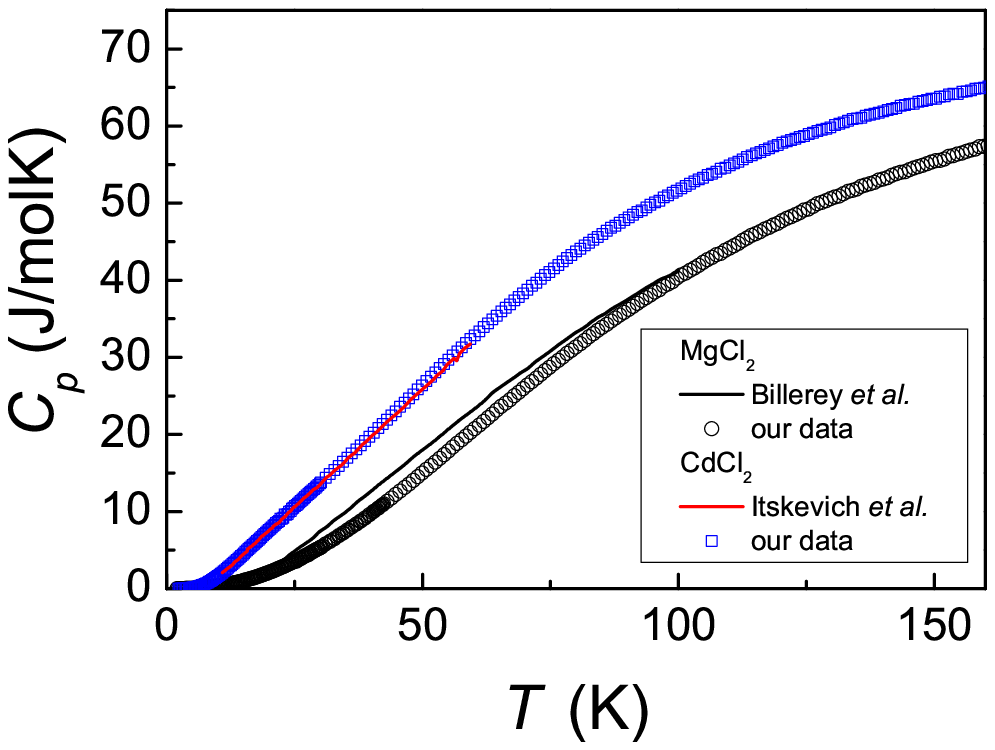}
\caption{\protect \label{LatticeCps}(Color online) Temperature dependence
of the specific heat of a single crystal of MgCl$_2$ (black $\circ$)
and of CdCl$_2$ (blue $\Box$) compared with literature  data as quoted in the inset (Refs. \onlinecite{Billerey1982,Itskevich1956}).}
\end{figure}

Additionally we determined the heat capacity of a small polycrystalline beads, typical diameter 2mm)  of CdCl$_2$ (Alfa Aesar 99.998\%)
The heat capacity of CdCl$_2$ has been measured before by Itskevich \textit{et al.}(Ref. \onlinecite{Itskevich1956}) in a very limited temperature range.  Takeda \textit{et al.} (Ref. \onlinecite{Takeda1984}) also determined the heat capacity of CdCl$_2$ in order to subtract the lattice contribution to the heat capacity of magnetic VCl$_2$,  but no data are available from the work of Takeda \textit{et al.}. Only the low-temperature cubic term $\propto T^3$ is quoted.
Figure \ref{LatticeCps} shows the heat capacities of MgCl$_2$ and CdCl$_2$ as determined in this work in comparison with literature data.

Figure \ref{LatticeCps} displays the heat capacities of CuCl$_2$ and
of MgCl$_2$ and CdCl$_2$ for comparison. As expected, the heat
capacity of MgCl$_2$ is significantly smaller, that of CdCl$_2$
markedly greater than that of CuCl$_2$. A detailed discussion how to construct an adequate lattice reference for CuCl$_2$ and the magnetic contributions to the heat capacity of CuCl$_2$ will be given in Section \ref{SectionExpvsTh}.

\section{Neutron Diffraction}

\subsection{Elastic Magnetic Scattering of a Powder Sample and  a Twinned Crystal}

\begin{figure}[ht]
\includegraphics[width=8cm]{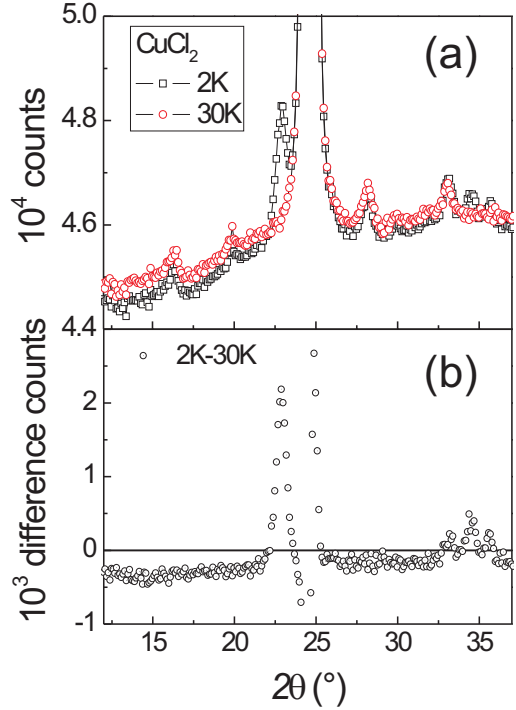}
\caption{\protect \label{D20diff}(Color online) Neighborhood of the
0\,0\,1 nuclear Bragg reflection of CuCl$_2$ ($\lambda$=2.4 \AA\ ) at
2 K and 30K. The additional magnetic Bragg reflection near 2$\Theta
\approx$ 23.2$^{\rm o}$ is clearly visible. It can be indexed as
-1,0\,225\,0.5.}
\end{figure}

In order to search for magnetic Bragg reflections of the long-range ordered AFM state we examined in detail the low-angular parts of the powder diffraction patterns.
A comparison of two diffraction patterns collected in D20's highest
intensity mode at a wavelength of $\lambda$=2.4 \AA\  at 2 K and 30K reveals
additional magnetic scattering up to 2$\Theta \sim$ 40$^{\rm o}$
(Figure \ref{D20diff}(a)). Most pronounced is a  reflection $\sim$
23$^{\rm o}$ at the low angle shoulder of the nuclear 0\,0\,1 reflection. Additional, though significantly less intense,
magnetic scattering intensity is visible in the 2 K pattern at $\sim$34$^{\rm o}$.
Both features are not present any more in the pattern collected
at 30K and thus can be clearly identified as magnetic Bragg
reflections.

Towards low scattering angles,  the 30K pattern lies clearly
above the 2 K pattern indicating significant paramagnetic scattering in the 30 K pattern.
In the difference pattern the magnetic Bragg
reflection near 19$^{\rm o}$ is slightly deformed owing to a small
shift of the 0\,0\,1 nuclear reflection arising from the contraction
of the lattice between 30 K and 2 K.

\begin{figure}
\includegraphics[width=8cm]{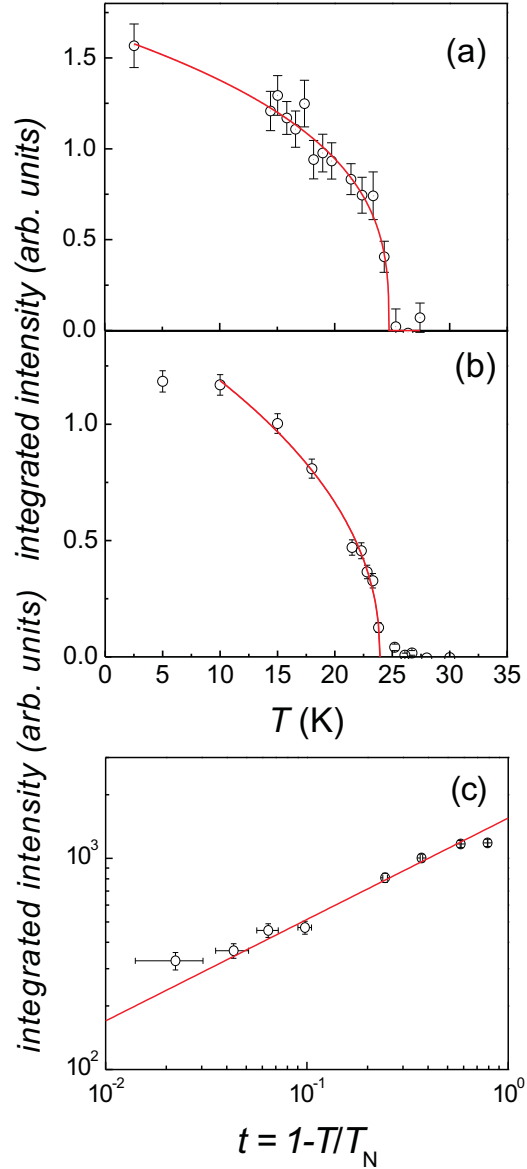}
\caption{\protect \label{Integ}(Color online) (a) Temperature dependence
of the integrated intensity of the magnetic Bragg reflection
observed at 2$\Theta \approx$ 18.2$^{\rm o}$ ($\lambda$=1.888 \AA\ )
($\circ$). The solid line is a fit of the experimental data with a
power law with a critical temperature $T_{\rm N}$=24.7(1)K.
(b) Temperature dependence
of the integrated intensity of the magnetic Bragg reflection
collected on a crystal of CuCl$_2$ (D10 four-circle diffractometer) $\lambda$=2.36 \AA\ )
($\circ$). The solid line is a fit of the experimental data with a
power law with a critical temperature $T_{\rm N}$=23.9(1)K and a critical exponent $\beta$=0.23(2).
(c) log-log plot of the integrated intensity displayed in (b) vs. the reduced temperature $t$=1-$T_{\rm N}/T$. The solid line represents a power law with a critical exponent $\beta$=0.24(2).}
\end{figure}

In order to obtain the temperature dependence of  this first
magnetic Bragg reflection, we collected a series of  patterns between 2 K and 45 K in the D20's higher resolution
mode using a wavelength of $\lambda$=1.888 \AA\. The difference
between the pattern collected at the respective temperatures and the
pattern at 28.9 K was fitted to a Gaussian the intensity of which is
displayed versus temperature in Figure  \ref{Integ}.

The temperature dependence of the integrated intensities can be well fitted by a critical power law according to Eq. (\ref{powlaw}) with  the critical exponent $\beta$

\begin{equation}\label{powlaw}
I(T)=I_0\,(1 - T/T_{\rm N})^{2\beta}.
\end{equation}

A fit of the integrated intensities converges to a N$\acute{\rm e}$el temperature $T_{\rm N}$=24.7(1) K consistent with the magnetic susceptibility and the heat capacity and a critical exponent $\beta$=0.16(1) (see Figure \ref{Integ}(a)).

Intensities of a set of magnetic Bragg reflections were collected on a twinned crystal.
Figure \ref{Integ}(b,c) display the intensity of a strong magnetic reflection collected on a twinned crystal on ILL's four-circle diffractometer D10.\cite{MBThesis,Rekretbp} Initially, this reflection was indexed as 0\, 0.776\,0.5. In view of the magnetic structure refinement of the powder data (see below) we must conclude that this reflection originates from the related twin individuum in the crystal.
Using the inverse of the twin matrix eq. (1) the indexes transform  to 0\,-0.776\,-0.5 i.e. 1\,1\,0 - $\vec{\tau}$ and 1\,1\,1 -  $\vec{\tau}$, where $\vec{\tau}$ is the magnetic propagation vector refined from the powder data below.
A fit of the integrated intensities of this reflection (solid line in Figure  \ref{Integ}(b,c)) gives $T_{\rm N}$=23.9(1) and $\beta$=0.23(1) consistent with the  powder diffraction results.

Although the investigated temperature regime is limited, and corrections for critical scattering near $T_{\rm N}$ have not been applied, the refined critical exponents $\beta$   obtained from the powder and crystal  are significantly lower than those characterizing the well known  three-dimensional universality classes (Ising, XY and Heisenberg) for which $\beta$  ranges between 0.325 and 0.365.\cite{Guillou1977}
The refined critical exponent $\beta \sim$ 0.2 for CuCl$_2$, however, comes close to the predictions of Kawamura for the universality class of of the SO(3) or the  Z$_2$ $\times$ S$_1$ symmetry for systems with chiral order. For such systems $\beta$  ranges between 0.22 and 0.25.\cite{Kawamura,Collins1989,Gaulin1994}
Figure \ref{Integ}(c) displays the intensity of a magnetic reflection  collected on a twinned crystal on ILL's four-circle diffractometer D10 together with a power law with a critical  exponent $\beta$ = 0.24(2)

The difference between the 2 K and the 30 K powder diffraction patterns (lower panel of Figure  \ref{D20diff})
reveals an additional weak triple of magnetic Bragg reflections between 32$^{\rm o}$
and 37$^{\rm o}$ with a peak shape very similar to those of neighboring
nuclear reflections.
Additionally, the difference patterns allows one to identify magnetic scattering around a scattering angle of $\sim$70$^{\rm o}$.

\begin{figure}[ht]
\includegraphics[width=8cm]{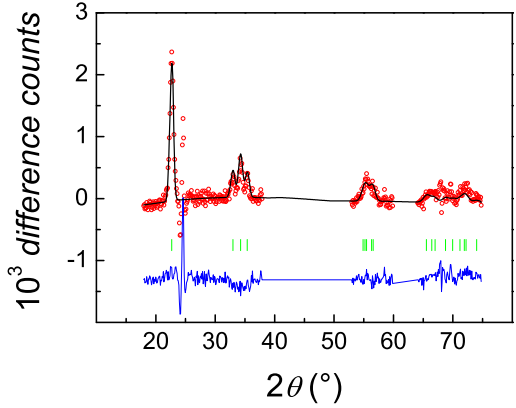}
\caption{\protect \label{Powmag}(Color online)Neutron diffraction
pattern (difference of 2 K and 30K patterns) (red $\circ$) ($\lambda$=2.4 \AA\ ) of a powder sample of
CuCl$_2$ collected  at the high intensity diffractometer D20
plotted in comparison with the pattern obtained from a profile
refinement assuming the magnetic structure described in detail in the text(solid black line). The difference is shown in
the lower part by the (blue) solid line. The positions of the the
magnetic Bragg reflections used to simulate the patterns are indicated by the
small (green) vertical bars in the lower part of the figure.}
\end{figure}

All magnetic Bragg reflections can be indexed with a propagation vector (1, 0.225,0.5) indicating a doubling of the magnetic cell along the \textit{c} axis and an incommensurate AFM ordering along \textit{b} (see Figure \ref{Powmag}).

The magnetic structure of CuCl$_2$ was refined using the D20 2 K and the difference powder diffraction data sets using the program Fullprof. In order to obtain reliable parameters for the nuclear structure (nuclear scale factor, profile parameters),  we first refined the 2 K  pattern assuming nuclear scattering only. Since magnetic scattering is very small this refinement converges well and provides the lattice and profile parameters and the scale factor.  Subsequently a difference pattern obtained by subtracting the 30 K pattern from the 2 K pattern was used to refine the magnetic structure on the basis of the propagation vector and the nuclear scale factor, which was kept fixed in the refinements. The background of the difference pattern was chosen manually and subtracted.
The refinement based on 18 magnetic reflections was successful assuming the nuclear space group $C \bar{\rm 1}$ and a helix  with the moments confined to the $bc$ plane.
Several tests were carried with the moment directions pointing in a more general direction out of the $bc$ plane   but did not lead to significantly improved fits.

Assuming this model for the magnetic structure the ordered moment at 2 K converged to

\begin{equation*}
\mu ({\rm 2 K}) = 0.50(1) \mu_{\rm B}.
\end{equation*}

\begin{figure}[ht]
\includegraphics[width=7cm,angle=90]{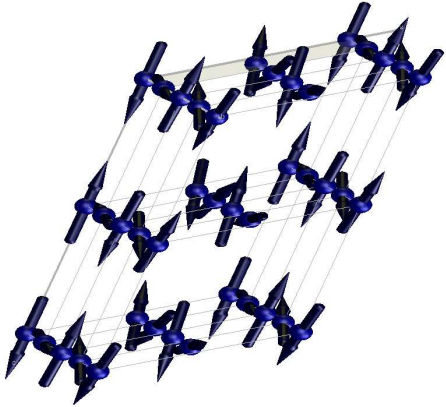}
\caption{\protect \label{MagStruc}(Color online) Magnetic structure of CuCl$_2$ at 2 K. A helix propagating along \textit{b} with the moments confined to the  $bc$ plane.
Neighboring moments along \textit{b} enclose an angle of $\sim$ 81$^{\rm o}$.}
\end{figure}

The propagation vector $\vec{\tau}$ at 2 K was refined to

\begin{equation*}
\vec{\tau} ({\rm 2 K}) = (1,0.2257(6),0.5)
\end{equation*}

 implying  an angle of 81.2(3)$^{\rm o}$ between neighboring moments along the chain. Figure \ref{MagStruc} displays the magnetic structure of CuCl$_2$ at 2 K.

\section{Density functional characterization of the spin exchange interactions and the spin-spiral state}

\begin{figure}
\includegraphics[width=6cm]{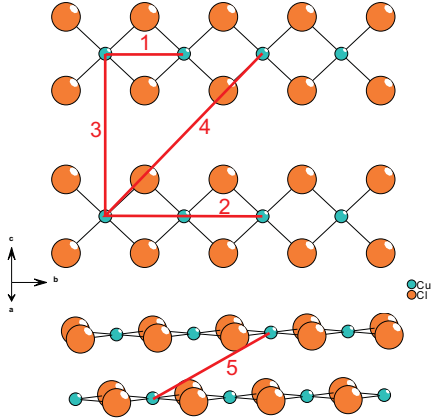}
\caption{\protect \label{Expaths}(Color online) Definition of the five exchange paths $J_1$ - $J_5$ of CuCl$_2$, where the numbers 1 - 5 represent $J_1$, ...  $J_5$, respectively. }
\end{figure}

To discuss the magnetic structure of CuCl$_2$, we consider the intra-chain and the inter-chain spin exchange interactions defined in Figure \ref{Expaths}. $J_1$ and $J_2$ are the NN and NNN intra-chain spin interactions, respectively. $J_3$ and $J_4$ are the inter-chain spin exchange interactions along the $c$-direction, and $J_5$ is that along the a-direction. The structural parameters associated with the spin exchange paths $J_1$ - $J_5$ are summarized in Table \ref{Table2}.

\begin{table}
\begin{ruledtabular}
\begin{tabular}{cccc}
  path& Cu...Cu (\AA\ )  &  Cl...Cl (\AA\ ) & $\angle$ Cu-Cl...Cl ($^{\rm o})$\\
\hline\\
$J_1$ & 3.299 & - & - \\ \\
$J_2$ & 6.599 & 3.299 ($\times$2) & 136.8\\ \\
$J_3$ &6.824 & 3.729 ($\times$2) & 133.1 \\ \\
$J_4$ &9.493 & 4.979 & 174.1 \\ \\
$J_5$ &7.608 & 3.808 & 131.7 \\
\end{tabular}
\end{ruledtabular}
\caption[]{Geometrical parameters associated with the spin exchange paths $J_1$ - $J_5$ of CuCl$_2$.
\label{Table2}}
\end{table}

\begin{figure}
\includegraphics[width=8cm]{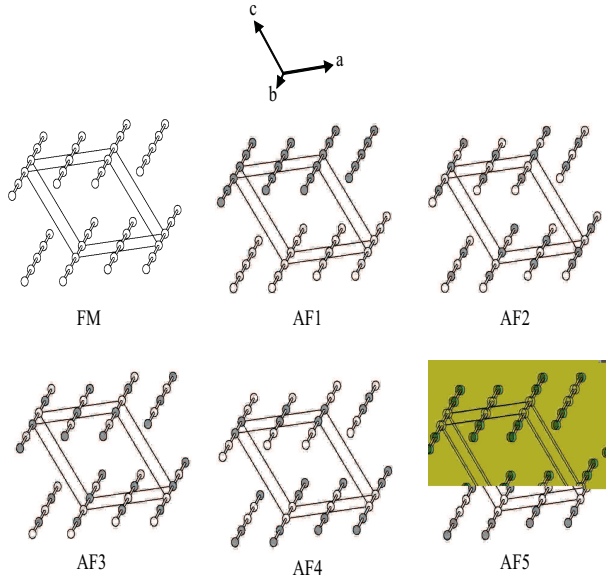}
\caption{\protect \label{Sixspin}(Color online) Six ordered spin arrangements of CuCl$_2$ employed to extract the spin exchange parameters $J_1$ - $J_5$. The up- and down-spin Cu sites are represented by the open and filled circles, respectively. }
\end{figure}

To estimate the spin exchange parameters $J_1$ - $J_5$, we first determine the total energies of the six ordered spin states of CuCl$_2$ presented in Figure \ref{Sixspin} on the basis of first principles density functional theory (DFT) electronic band structure calculations. Our DFT calculations employed the Vienna ab initio simulation package (Ref. \onlinecite{Kresse1993,Kresse1996a,Kresse1996b}) with the projected augmented-wave method, the generalized gradient approximation (GGA) for the exchange and correlation functional, (Ref. \onlinecite{Perdew1996}) the plane-wave cut-off energy of 330 eV, and the sampling of the irreducible Brillouin zone with 36 $k$-points. To take into consideration of the strong electron correlation associated the Cu 3d state, we performed GGA plus onsite repulsion (GGA+U) calculations (Ref. \onlinecite{Dudarev1998}) with $U$ = 3, 5 and 7 eV on Cu. The relative energies of the six ordered spin states obtained from our GGA+U calculations are summarized in Table \ref{Table3}.

\begin{table}
\begin{ruledtabular}
\begin{tabular}{cccc}
  state & $U$ = 3eV &  $U$ = 5eV & $U$ = 7 eV\\
\hline\\
FM & 0 & 0  & 0  \\ \\
AFM1 & -11.23 & -8.62  & -6.52  \\ \\
AFM2 & 21.68 & 20.06  & 17.43  \\ \\
AFM3 & 13.77 & 14.09  & 12.95  \\ \\
AFM4 & -31.01 & -22.12  & -15.93  \\ \\
AFM5 & -39.72 & -28.85  & -21.05  \\
\end{tabular}
\end{ruledtabular}
\caption[]{Relative energies (in meV) of the five ordered spin states of CuCl$_2$ obatined from the GGA+U calculations.
\label{Table3}}
\end{table}

The energies of the six ordered spin states can also be written in terms of the spin Hamiltonian,

\begin{equation}
\label{Hamilton}
H = -\sum J_{ij} \vec{{S_i}}\vec{{S_j}},
\end{equation}
							(1)
where $J_{\rm ij}$ (= $J_1$ - $J_5$) is the spin exchange parameter for the spin exchange interaction between the spin sites $i$ and $j$, while  $\vec{{S_i}}$ and $\vec{{S_j}}$  are the spin angular momentum operators at the spin sites $i$ and $j$, respectively. By applying the energy expressions obtained for spin dimers with N unpaired spins per spin site (in the present case, N = 1), (Ref. \onlinecite{Dai2001,Dai2003}) the total spin exchange energies per formula unit of the six ordered spin states (see Figure  \ref{Sixspin}) are written as

\begin{eqnarray}
E_{FM} & = & (-2J_1 - 2J_2 - 2J_3 - 4J_4 - 4J_5)\frac{N^2}{4}\nonumber\\
E_{AF1}& = & (-2J_1 - 2J_2 + 2J_3 + 4J_4 + 4J_5)\frac{N^2}{4}\nonumber\\
E_{AF2}& = & (+2J_1 - 2J_2 - 2J_3 - 4J_4 + 4J_5)\frac{N^2}{4}\nonumber\\
E_{AF3}& = & (+2J_1 - 2J_2 + 2J_3 + 4J_4 - 4J_5)\frac{N^2}{4}\nonumber\\
E_{AF4}& = & (+2J_1 - 2J_3 + 4J_4)\frac{N^2}{4}\nonumber\\
E_{AF5}& = & (+2J_1 + 2J_3 - 4J_4)\frac{N^2}{4}\nonumber\\
\end{eqnarray}

\begin{table}[tbh]
\begin{ruledtabular}
\begin{tabular}{cccc}
  state & $U$ = 3eV &  $U$ = 5eV & $U$ = 7 eV\\
\hline\\
$J_1$ & 23.3 & 21.4  & 18.4  \\ \\
$J_2$ & -41.4 & -31.9  & -24.5  \\ \\
$J_3$ & -9.1 & -7.0  & -5.3  \\ \\
$J_4$ & -0.2 & -0.1  & -0.1  \\ \\
$J_5$ & -0.8 & -0.7  & -0.5  \\
\end{tabular}
\end{ruledtabular}
\caption[]{Values of the spin exchange parameters $J_1$ - $J_5$ (in meV) in CuCl$_2$ obtained from the GGA+U calculations.
\label{Table4}}
\end{table}

By mapping the relative energies of the six ordered spin states given in terms of the spin exchange parameters to the corresponding relative energies obtained from the GGA+U calculations, we obtain the values of the spin exchange parameters $J_1$ - $J_5$ summarized in Table \ref{Table4}.

For all values of $U$ employed, the two strongest spin exchange interactions are the intra-chain interactions $J_1$ and $J_2$. As found for the magnetic oxides with CuO$_2$ ribbon chains, (Ref. \onlinecite{Xiang2007a,Xiang2007b}) $J_1$ is ferromagnetic while $J_2$ is antiferromagnetic with $J_2$ larger in magnitude than $J_1$. As a consequence, the intra-chain spin exchange interactions are geometrically frustrated. Except for $J_3$, the inter-chain spin exchange interactions are negligible; $J_3$ is antiferromagnetic and is weaker than $J_2$ by a factor of approximately four. Because of the intra-chain spin frustration, the magnetic ground state of each CuCl$_2$ chain is expected to be a spin-spiral state with a certain incommensurate repeat vector $q_y$ along the chain direction, i.e., the $b$-direction. The inter-chain spin arrangement is expected to be antiferromagnetic along the $c$-direction due to the spin exchange $J_3$. Our calculations of the energy E (0, $q_y$, 1/2) as a function of the modulation wave vector $\vec{q}$ = (0, $q_y$, 1/2) shows a minimum at $q_y \approx$ 0.22.

	The magnetic state with the spin-spiral arrangement removes inversion symmetry and hence should induce ferroelectric polarization. To confirm this prediction, we carried out GGA+U plus spin-orbit coupling (SOC) (Ref. \onlinecite{Kunes2001})  calculations for CuCl$_2$ that has the commensurate spin-spiral arrangement with $q$ = (0, 0.25, 0) to reduce the computational task. Indeed, our calculations of the ferroelectric polarization using the Berry phase method (Ref. \onlinecite{King1993,Resta1994})  for the spin-spiral state of CuCl$_2$ determined by the GGA+U+SOC calculations lead to nonzero ferroelectric polarizations; with the spin-spiral plane parallel ($\parallel$) and perpendicular ($\perp$ ) to the CuCl$_4$ ribbon plane, the $x$-, $y$- and $z$- components of the polarizations are calculated to be $P_{\parallel}$ = (-44.5, 0, 71.6) and $P_{\perp}$  = ( 5.8, 0, 28.3) in units of  $\mu$C/m$^2$.

\section{Experiment versus Theoretical Results}\label{SectionExpvsTh}

In the following we will compare the implications of the theoretical results for the magnetic susceptibility and the thermal properties. The calculations indicate that the intra-chain exchange is dominant and that the magnetic properties of CuCl$_2$ have to be described  as those of a $S$=1/2 frustrated Heisenberg chain with FM nearest and AFM NN interaction. According to the calculations, the ratio $J_1$/$J_2$ is  negative and is expected to be of the order of $\sim$ -0.5  to -0.75. Additionally, there is small but noticeable inter-chain interaction being dominant along the $c$ axis that leads to cooperative AFM ordering and the AFM arrangement along the $c$ direction with the doubling of the magnetic unit cell along $c$.  Magnetic and thermal properties of $S$=1/2 zig-zag Heisenberg chains have been calculated in great detail by exact diagonalization by Heidrich-Meisner and coworkers,  and numerical tables are available for a wide range of the ratio $\alpha = J_{\rm 1}$/$J_{\rm 2}$. \cite{Heidrich2006,Heidrichweb} These works comprise also the magnetic susceptibility of a $S$=1/2 Heisenberg chain with NN AFM exchange only, i.e., $\alpha$=0 as calculated before by Kl\"umper \textit{et al.} and parametrized by a Pad$\rm \acute{e}$ approximant by Johnston \textit{et al.}.\cite{Kluemper2000,Johnston2000}

\subsection{Magnetic Susceptibility}

Correcting  the magnetic susceptibility results with the sum of the temperature independent diamagnetic and Van Vleck corrections  to the  susceptibility ($\chi_{\rm 0}$) gives the pure spin susceptibility $\chi_{\rm spin}(T)$ of CuCl$_2$. Using the $g$-factor along $b$ as determined by the  EPR measurements, $g_b$=2.050(1), enables us to compare the magnetic susceptibility of a CuCl$_2$ crystal measured with magnetic field along the $b$-axis with model calculations. The only remaining adjustable parameters are the intra-chain exchange constants $J_{\rm 1}$ and $J_{\rm 2}$.
The effect of  the finite inter-chain exchange $J_{\rm inter}$ on the paramagnetic susceptibilities calculated for the zig-zag chains has been taken into account by employing a mean field approach, which gives as inter-chain interaction corrected susceptibility as(Ref. \onlinecite{Carlin1986})

\begin{equation}
\label{meanfield}
\chi_{\rm spin}^{\rm cor}(T) = \frac{\chi_{\rm spin}(T)}{1-(zJ/N_{\rm A}g_{\rm \perp}^2\mu_{\rm B}^2)\,\chi_{\rm spin}(T)}.
\end{equation}

The leading intra-chain exchange  $J_{\rm \perp}$ is provided by $J_{\rm 3}$ (see Table \ref{Table3} and Figure \ref{Integ})
and amounts to about 20\% of the next-nearest exchange $J_{\rm 2}$ along the chain. Including intra-chain exchange according to Eq. \ref{meanfield} renormalizes the calculated susceptibilities by about 8\% near the maximum at $\sim$73 K. At room temperature the decrease amounts to $\sim$4\%.

\begin{figure}
\includegraphics[width=8cm]{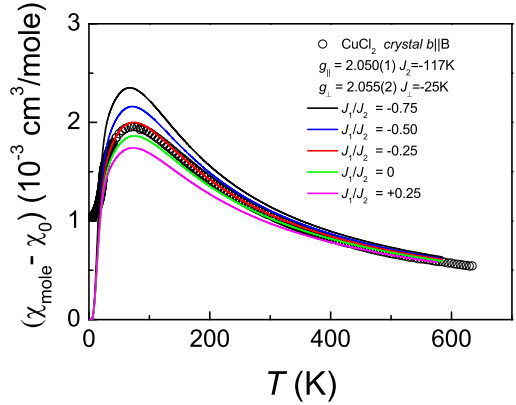}
\caption{\protect \label{xtalchi}(Color online) Temperature dependence
of the magnetic susceptibility  of a crystal of CuCl$_2$ measured with magnetic field aligned along the crystal $b$-axis corrected for the sum of diamagnetic and Van Vleck contributions ($\chi_{\rm 0}$) ($\circ$) compared with the results of  exact diagonalization calculations of $S$=1/2 Heisenberg chains with NN and NNN exchange $J_{\rm 1}$ and $J_{\rm 2}$, respectively defining the ratios  $\alpha$=$J_{\rm 1}$/$J_{\rm 2}$ as given in the inset. The (colored) solid lines were calculated according to Eq. \ref{meanfield} with $\chi_{\rm spin}(T)$ taken from Ref. \onlinecite{Heidrich2006,Heidrichweb}  for $J_{\rm 2}$= -117 K and $\alpha$s as given in the inset. From top to bottom $\alpha$ amounts to -0.75, ..., +0.25, respectively.}
\end{figure}

Figure \ref{xtalchi} shows the comparison of the spin susceptibility of CuCl$_2$ as determined in this work with the inter-chain exchange corrected susceptibility of a Heisenberg chain with NN and NNN exchange interaction $J_{\rm 1}$ and $J_{\rm 2}$, respectively.\cite{Heidrich2006,Heidrichweb}

According to this comparison the magnetic susceptibility of CuCl$_2$ is best reproduced by an inter-chain exchange corrected susceptibility of a Heisenberg chain with NNN exchange interaction $J_{\rm 2}$=-117 K and a ratio $\alpha\approx$ -0.25, i.e., ferromagnetic nearest neighbor exchange as predicted by the DFT calculations. However, the ratio $\alpha$ is markedly smaller than predicted by the DFT calculations (see Table \ref{Table3}). Including inter-chain exchange interaction is essential to reproduce the measured susceptibility correctly.

\subsection{Magnetic Heat Capacity}

The magnetic contributions to the heat capacity of CuCl$_2$ has been evaluated before by Stout and Chisholm and subsequently by Billerey and coworkers. \cite{Stout1962,Billerey1982}
Our measurements confirm (see Figure \ref{heat}(a)) their peak-anomaly at $\sim$23 K indicating  long-range AFM ordering. Our data agree very well with those of the previous works,
and the anomaly is somewhat sharper in our experiment  possibly due to the fact that polycrystalline samples have been used before.\cite{Stout1962,Billerey1982}
Stout and Chishom and Billerey \textit{et al.}
used the heat capacities of MnCl$_2$ and MgCl$_2$, respectively, in order
to construct an estimate of the lattice heat capacity of CuCl$_2$.
We redetermined the heat capacities of MgCl$_2$ and found marked deviations of our data from those of Billerey. We considered also CdCl$_2$ as another possible system to estimate the phonon contributions to the heat capacity of CuCl$_2$ and measured the heat capacity of CdCl$_2$ over a temperature range much wider than covered by the experiments of Itskevich \textit{et al.}(see Figure \ref{LatticeCps}(a)).
MgCl$_2$ and CdCl$_2$ are diamagnetic and have a crystal
structure  containing MgCl$_2$ and CdCl$_2$ layers, which closely
resemble those in the crystal structure of CuCl$_2$. However,
the MCl$_6$ (M=Mg, Cd) octahedra are not axially elongated as they are in the CuO$_6$ octahedra. In addition, owing to the different
atomic masses of Cu and Mg, the phonon spectrum of CuCl$_2$ and
MgCl$_2$ and Cd$_2$ may be expected to be different, especially at low energies. Although MnCl$_2$ is magnetic, its AFM ordering occurs at very low temperature, 1.81 K and 1.96 K,  and magnetic contributions to its heat capacity are negligible above $\sim$10K.\cite{Murray1955,Chisholm1962}
Above 10K, MnCl$_2$ may therefore provide a reasonable approximation to the lattice heat capacity of CuCl$_2$.
In order to compensate for the difference in the cation masses,  we employed the method of corresponding states and assume that the phonon spectra of MgCl$_2$, CdCl$_2$ and MnCl$_2$ are very similar to that of CuCl$_2$ and the different phonon energies can be compensated for by modifying the temperatures.\cite{Boo1977}
Compensation of the different cation mass was done by stretching or compressing the temperature scale by a constant factor. This factor was chosen such that at sufficiently high temperatures where magnetic contributions to the heat capacity vanish the heat capacities of the lattice references match with those of CuCl$_2$ (see Figure \ref{heat}). We found factors of 0.88, 0.93 and 1.18 to be appropriate for MgCl$_2$, MnCl$_2$ and CdCl$_2$, respectively.

A comparison of the heat capacities of CuCl$_2$ and MCl$_2$ (M=Mg, Mn, Cd) reveals significant short-range ordering magnetic contributions to the heat capacity of CuCl$_2$ up to $\sim$ 100 K, as already concluded by Stout \textit{et al.} and Billerey \cite{Stout1962,Billerey1982}.

In order to construct a lattice heat capacity reference for CuCl$_2$ from the heat capacities of MgCl$_2$, MnCl$_2$ and CdCl$_2$, we tried various combinations of the heat capacities of MgCl$_2$, MnCl$_2$ and CdCl$_2$ with temperatures scaled by the factors given above and subtracted these from the total heat capacity of CuCl$_2$.
The resulting magnetic heat capacity of CuCl$_2$, $C_{\rm mag}$/$T$ was subsequently integrated and the total magnetic entropy was compared with $R$ln2, the entropy expected for a $S$=1/2 magnetic system as a crosscheck.

Due to the very large deviations of the heat capacity of MgCl$_2$ from that of CuCl$_2$, especially
in the temperature regime below $\sim$ 50K where magnetic contributions are essential we discarded  MgCl$_2$  and achieved
an adequate lattice heat capacity reference  by  averaging for each temperature the heat capacities of MnCl$_2$ and CdCl$_2$.

\begin{figure}[ht]
\includegraphics[width=9cm]{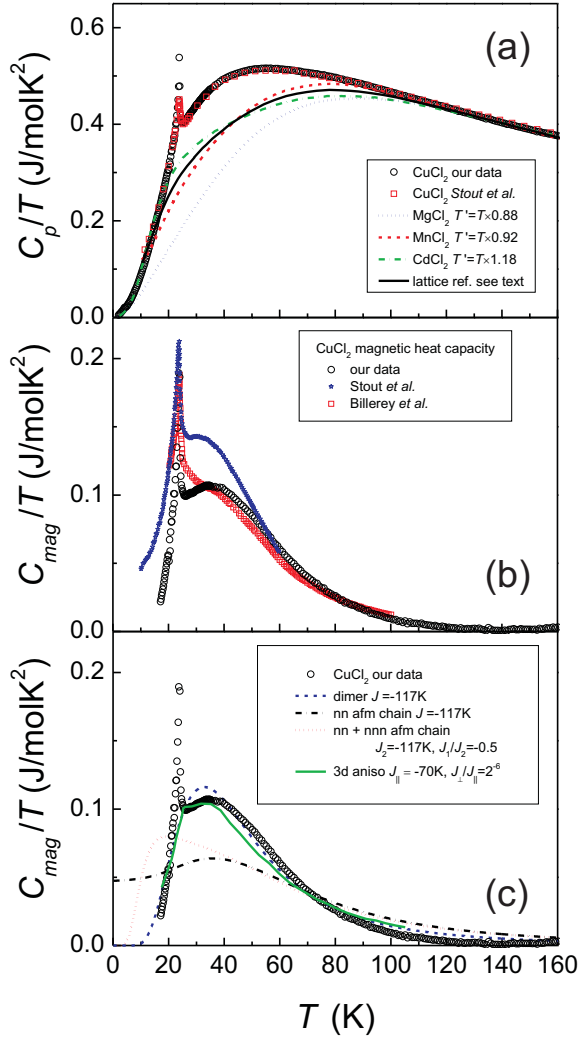}
\caption{\protect \label{heatphonon}(Color online) (a) Temperature dependence
of the specific heat of a single crystal of CuCl$_2$ (black $\circ$) and of a polycrystalline sample measured by Billerey \textit{et al.} (red $\Box$). \cite{Billerey1982}
The heat capacities of MgCl$_2$ and CdCl$_2$ as measured in this work and of MnCl$_2$ (Ref. \onlinecite{Chisholm1962}) are also given. Their temperature axes have been rescaled by the factor 0.88, 0.92 and 1.18 for MgCl$_2$, MnCl$_2$ and CdCl$_2$, respectively.
The (black) solid line is the lattice reference (phonon contribution to the heat capacity) constructed from the heat capacities of MnCl$_2$ and CdCl$_2$ as described in detail in the text.
(b) Magnetic contribution to the heat capacity of CuCl$_2$ obtained as the difference of the CuCl$_2$ data and the lattice reference displayed in (a).
Published data by Stout \textit{et al.} and Billerey \textit{et al.} are displayed for comparison.
(c) Magnetic heat capacity of CuCl$_2$ (black $\circ$) compared with various model calculations as labeled in the inset and discussed in detail in the text.}
\end{figure}

The magnetic heat capacity of CuCl$_2$ obtained after subtracting the phonon contributions is shown in Figure \ref{heatphonon}(b) and (c).
In addition to the $\lambda$-type anomaly at $\sim$23 K the  magnetic contributions to the heat capacity are characterized by a broad anomaly due to short-range AFM ordering with maximum at 35 K (in $C_{\rm mag}/T$ and a maximum value of 0.11 J/molK$^2$).
Our data are close to those of Billerey \textit{et al.} but deviate markedly from the data published by Stout \textit{et al.}, who found a significantly higher value for the short-range ordering maximum. The total entropy of the magnetic heat capacity amounts to 0.66(2)$R$, very close to  $R$ln2 expected for the entropy of a spin system with $S$=1/2. Figure \ref{heatphonon}(c) shows a comparison of the magnetic heat capacity with various model calculations, e.g., the heat capacity of an AFM Heisenberg chain with NN only and with NN and NNN interactions with exchange parameters $J_{\rm 1}$ and the ratio $J_{\rm 2}$/$J_{\rm 1}$ similar to those discussed for the magnetic susceptibility.\cite{Johnston2000,Heidrich2006}
Both chain models deviate significantly from the experimental data.
Surprisingly good agreement of the broad magnetic heat capacity anomaly is obtained for a simple model assuming a spin $S$=1/2 dimer according to

\begin{equation}
\label{Hamilton}
H = - J_{\rm dim} \vec{{S_1}}\vec{{S_2}},
\end{equation}

with the heat capacity discussed in detail in Ref. \onlinecite{Johnston2000}
for an exchange constant $J_{\rm dim} \approx$ -117 K. The data are also well reproduced by an anisotropic three-dimensional Heisenberg model on a cubic lattice with in-plane exchange $J_{\rm \parallel}$ and interplane coupling $J_{\rm \perp}$.\cite{Sengupta2003} With $J_{\rm \parallel} \approx$ -70K and  $J_{\rm \perp} \approx 2^{-6} J_{\rm \parallel}$ maximum value and position of the short-range ordering anomaly are rather well reproduced. The onset of three-dimensional ordering showing up as a well separated peak well below the short-range ordering maximum appears close to the cooperative ordering temperature observed for CuCl$_2$.

\section{Discussion}

Complex compounds containing Cu$^{2+}$ ions with low-dimensional magnetic behavior are legion and have been widely investigated.\cite{DeJongh1974}
However, the detailed magnetic behavior of anhydrous Copper(II)-chloride, CuCl$_2$, itself, however,  has largely remained a mystery until now.
From the early susceptibility studies by deHaas and Gorter and Starr \textit{et al.} it had become evident that CuCl$_2$ shows unusual AFM properties with a broad maximum in the magnetic susceptibility.\cite{DeHaas1931,Starr1940}
The specific heat of CuCl$_2$ measured by Stout and Chisholm proved long-range AFM ordering at around 23 K and the excess heat capacity extending well above $T_{\rm N}$ was analyzed in terms of linear chain behavior. Because of a lack of better theoretical foundation this analysis was done using an Ising-type approach.\cite{Stout1962}  A neutron diffraction study  in order to determine the long-range ordered groundstate  has not been carried out so far.

Our study presents a detailed confirmation of the early susceptibility study carried out on polycrystalline samples and twinned single-crystals of anhydrous CuCl$_2$. Susceptibility and heat capacity confirm the one-dimensional AFM properties of CuCl$_2$ and  the N$\acute{\rm e}$el temperature 23 K as already found by Stout and Chisholm.\cite{Stout1962}
This rather high N$\acute{\rm e}$el temperature as compared to the maximum temperature of the short-range ordering susceptibility maximum occurring at about 70K indicates substantial inter-chain exchange interaction and underlines the conclusion  by deJongh and Miedema that CuCl$_2$ is a rather poor example of a chain structure. \cite{DeJongh1974}

CuCl$_2$ consists of  ribbons of edge-sharing CuCl$_4$ square planes with Cu$^{2+}$ ions. In such systems a spin helicoidal groundstate can set in along the chain direction at low temperature due to the competition between NN ferromagnetic and NNN antiferromagnetic interactions along the Cu$^{2+}$ chains. The first clear example proven to show such a behavior was LiCuVO$_4$ that crystallizes with an inverse spinel structure and contains CuO$_2$ ribbon chains.
Helicoidal magnetic ordering with the helix propagating along the chain direction was found by a single-crystal neutron diffraction study.\cite{Gibson2004} The frustration scenario with ferromagnetic NN and larger AFM NNN exchange was proven by an inelastic neutron scattering study.\cite{Enderle2005} Special interest in the magnetic properties of LiCuVO$_4$ arose lately from the observation of multiferroic behavior found by Naito \textit{et al.} and Schrettle \textit{et al.}.\cite{Naito2007,Yasui2008,Schrettle2008} LiCuVO$_4$ shows a dielectric polarization below the N$\acute{\rm e}$el temperature and the polarization can favorably be switched by applying a magnetic field. The origin of this dielectric polarization is due to spin-orbit coupling on the Cu sites, however, the asymmetric charge density distribution  necessary for the electric polarization occurs around the O$^{2-}$ ions.\cite{Xiang2007a}

LiCu$_2$O$_2$ is another Cu$^{2+}$ system with CuO$_2$ ribbon chains exhibiting helicoidal spin order.
\cite{Masuda2004,Masuda2005,Gippius2004,Drechsler2005,Papagno2006,Mihaly2006,Xiang2007a}
In LiCu$_2$O$_2$, in contrast to LiCuVO$_4$, the magnitude of the  FM NN (-11$\pm$3 meV) exchange is larger than the AFM NNN exchange (7$\pm$1 meV). Ferroelectricity in LiCu$_2$O$_2$ has been shown by Park and coworkers.\cite{Park2007}

Another related system  isostructural with LiCu$_2$O$_2$, which lately has attracted special attention, is NaCu$_2$O$_2$.  A helicoidal magnetic ground state has been found by neutron powder diffraction and by NMR experiments.\cite{Capogna2005,Choi2006,Drechsler2006,Gippius2008}
A review of recent systems with helicoidal ground states  has been compiled by Drechsler and collaborators.\cite{Drechsler2007}

All helicoidal systems investigated so far contain oxygen ions to provide superexchange between the Cu$^{2+}$  moments. Clear proof of the helicoidal ground state by magnetic neutron scattering is available only for a few of them. Evidence for systems containing anions other than O$^{2+}$ has not been reported yet.

Anhydrous CuCl$_2$ is  therefore the first example for a halide system for which a helicoidal ground state has been proven unequivocally.  Our electronic structure calculations give clear theoretical evidence for a  scenario of competing FM NN and AFM NNN exchange interactions.
Our experimental results for the bulk magnetic properties and the comparison with their predictions for a $S$=1/2 AFM Heisenberg chain with competing FM NN and AFM NNN exchange support the theoretical predictions. Clear proof for the helicoidal AFM ground state comes from the neutron diffraction work on powder and single crystals. We find a magnetic structure indicating a $S$=1/2 helix in the CuCl$_2$ ribbon chains. The component of the magnetic propagation vector along the $b$-axis
amounts to 0.2257(6) and is in best agreement with the result of the electronic structure calculations. The spin-spiral as refined from diffraction data to is found to lie in the $bc$ plane, consistent with the  spin-flop observed in the magnetization data when the magnetic field is oriented along the $b$ axis.
The refined magnetic structure indicates a magnetic unit cell which is close to being quadrupled as compared to the nuclear cell with neighboring moments enclosing an angle of $\sim$81$^{\rm o}$ which leads to a significant reduction of NN exchange energy. With respect to the helix this scenario is analogous to that in LiCuV$_4$ (propagation vector (0,0.468,0) (Ref. \onlinecite{Propvector}) and NaCu$_2$O$_2$ (propagation vector (0.5,0.227,0). \cite{Capogna2005}

The magnetic susceptibility data indicate the importance of inter-chain interaction, which according to the electronic structure calculation amounts to $\sim$20\% - 30\% of the intra-chain interaction. Interchain interaction decreases the temperature for  the short-range order maximum and provides a better agreement of experimental data and theoretical prediction (Figure \ref{xtalchi}).

Particularly surprising and striking are the deviations of the short-range ordering contributions to the magnetic heat capacity for CuCl$_2$ and the heat capacity expected for a $S$=1/2 quantum chain with FM NN and AFM NNN exchange interactions (see Fig. \ref{heatphonon}(c)). The short-range order contributions match significantly better with a $S$=1/2 dimer singlet-triplet excitation scenario with AFM exchange $\sim$ -117 K corresponding to the exchange found for the NNN interactions along the ribbons. We believe that this coincidence is due to the competing intra-chain competing interactions and the additional AFM inter-chain coupling.
The magnetic configuration energy $U_{\rm mag}$ measures the short-range pair correlation function  according to (cf. e.g. Ref. \onlinecite{Fisher1962})

\begin{equation}
U_{\rm mag} \propto J \langle \vec{{S_0}}\,\vec{{S_1}}\rangle,
\end{equation}

where the subscript 1 denotes a spin adjacent to 0. Thus  $U_{\rm mag}$ and the magnetic heat capacity $dU_{\rm mag}/dT$ in a frustrated chain system with coupling to neigboring chains may primarily sense magnetic correlations in a short-ranged cluster in which the configuration energy is determined by the leading AFM exchange to NNN spins.

In summary, anhydrous CuCl$_2$ shows 1D AFM behavior and long range AFM ordering below a N$\acute{\rm e}$el temperature of 23.9 K, below which CuCl$_2$ adopts an incommensurate magnetic structure (1,0.2257,0.5) with a spin-spiral propagating along $b$ and the moments confined in the $bc$ crystallographic plane. The spin-spiral results from competing FM NN and AFM NNN spin exchange interactions along $b$. Anhydrous CuCl$_2$ is the first halide quantum system containing CuCl$_2$ ribbon chains for which a helicoidal magnetic ground state is realized, and is expected to be ferroelectric below 23.9 K.

\begin{acknowledgments}
We thank E. Br\"ucher, S. H\"ohn, S. Lacher and G. Siegle for
experimental assistance.
The work at North Carolina State University was supported by the Office of Basic Energy Sciences, Division of Materials Sciences, U. S. Department of Energy, under Grant DE-FG02-86ER45259.
\end{acknowledgments}

\bibliographystyle{apsrev}

\begin{thebibliography}{99}
\bibitem{Propvector}
Note that in LiCuV$_4$ the nuclear cell contains two Cu atoms along $b$ due to the VO$_4$-tetrahedra alternating in their orientation. Accordingly,  the propagation vector (0,0.468,0) indicates an approximate doubling of the magnetic cell which again contains four $S$=1/2 entities.
\end{thebibliography}

\end{document}